\documentclass[aps,prl,reprint,superscriptaddress]{revtex4-1}

\usepackage[utf8]{inputenc}

\usepackage{ifpdf}

\pdfoutput=1 

\ifpdf
\usepackage[pdftex]{graphicx}
\else
\usepackage{graphicx}
\fi

\begin{document}

\title{High fidelity single-shot readout of a transmon qubit using a SLUG $\mu$wave amplifier}
\date{\today}
\author{Yanbing Liu}
\affiliation{Department of Electrical Engineering, Princeton University, Princeton, New Jersey 08544, USA}
\author{Srikanth Srinivasan}
\affiliation{Department of Electrical Engineering, Princeton University, Princeton, New Jersey 08544, USA}
\affiliation{IBM T. J. Watson Research Center, Yorktown Heights, New York 10598, USA}
\author{D. Hover}
\affiliation{Department of Physics, University of Wisconsin, Madison, Wisconsin 53706, USA}
\author{Shaojiang Zhu}
\affiliation{Department of Physics, University of Wisconsin, Madison, Wisconsin 53706, USA}
\author{R. McDermott}
\affiliation{Department of Physics, University of Wisconsin, Madison, Wisconsin 53706, USA}
\author{A. A. Houck}
\affiliation{Department of Electrical Engineering, Princeton University, Princeton, New Jersey 08544, USA}

\ifpdf
\DeclareGraphicsExtensions{.pdf, .jpg, .tif}
\else
\DeclareGraphicsExtensions{.eps, .jpg}
\fi

\begin{abstract}
We report high-fidelity, quantum nondemolition, single-shot readout of a superconducting transmon qubit using a DC-biased superconducting low-inductance undulatory galvanometer(SLUG) amplifier. 
The SLUG improves the system signal-to-noise ratio by $7~\mathrm{dB}$ in a $20~\mathrm{MHz}$ window compared with a bare HEMT amplifier.
An optimal cavity drive pulse is chosen using a genetic search algorithm, leading to a maximum combined readout and preparation fidelity of $91.9\%$ with a measurement time of $T_{\mathrm{meas}} = 200 \mathrm{ns}$. Using post-selection to remove preparation errors caused by heating, we realize a combined preparation and readout fidelity of $94.3\%$. 
\end{abstract}

\maketitle
Scalable fault-tolerant quantum computation with superconducting qubits requires high fidelity measurement. Moreover, having a quantum non-demolition (QND) measurement, in which a measured qubit remains in its state after measurement, enables or facilitates many quantum information processing techniques such as state preparation through measurement and error correction.  Long coherence times and low noise amplifiers are both essential in resolving a qubit state quickly and with high-fidelity.
Several kinds of Josephson parametric amplifiers (JPA) have been shown to operate at or near the quantum limit, and these indeed enabled high fidelity($>93\%$) qubit measurement~\cite{Johnson2012,Riste2012,Hatridge2013}. Similarly, microstrip SQUID amplifiers (MSA)~\cite{Johnson2011,Hoffman2011} have been used to reduce measurement noise in cQED, but it is still challenging to engineer the MSA so that it has large enough gain and low enough noise at relevant microwave frequencies~\cite{Spietz2008, DeFeo2010, DeFeo2012}. 

In this letter, we rely on a SLUG~\cite{Ribeill2011} microwave amplifier to boost the measurement signal-to-noise ratio (SNR). 
The SLUG is incorporated as a preamplifier preceding the standard high electron mobility transistor (HEMT) amplifier. Unlike parametric amplifiers which require a microwave pump, the SLUG needs only two DC biases for current and flux through its SQUID loop. To sample gigahertz oscillations, the input microwave current is directly injected into the DC-SQUID loop. The resulting oscillatory output voltage serves as an amplified signal. Highly optimized devices are expected to achieve gain greater than $15~\mathrm{dB}$, bandwidth of several hundred MHz, and added noise of order one quantum in the frequency range of 
$5-10~\mathrm{GHz}$~\cite{Hover2012}. 
In contrast to existing parametric amplifiers, the SLUG has a maximum input power of tens of photons and it does not require a strong microwave pump tone.

\begin{figure}
\centering
\includegraphics[scale=1]{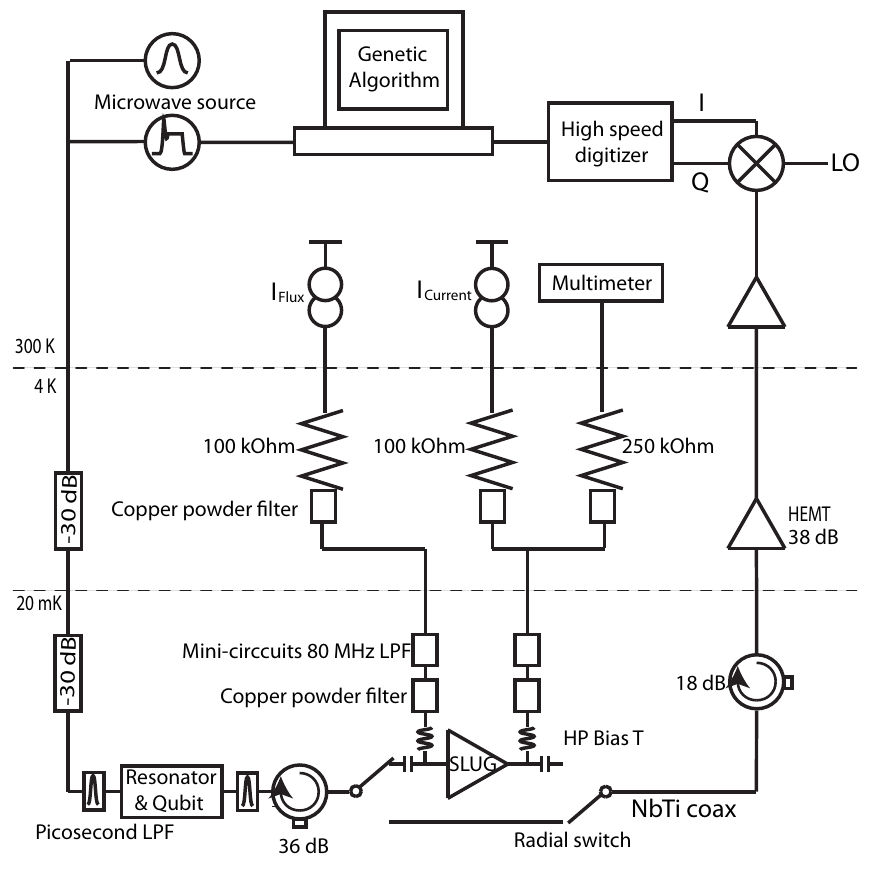}
\caption{Schematic representation of the measurement setup. Both the device and the SLUG amplifier are placed in $\mu$-metal magnetic shields. The SLUG is DC biased through two bias T.}
{\label{fig:Setup}}
\end{figure}

In our cQED system, we have a transmon qubit dispersively coupled to a coplanar waveguide $\lambda/2$ resonator. The system then is well-described by Jaynes-Cummings model in the linear regime. The measurement exploits the fact that the cavity resonance frequency depends on the state of the qubit~\cite{Schuster2005}. In other words, the qubit information is encoded in the amplitude and phase of microwave transmission through the cavity. The device is mounted in a copper box wrapped with MCS ECCOSORB tape (Emerson and Cuming) to protect it from external radiation and anchored to the $20~\mathrm{mK}$ base temperature stage of a dry dilution refrigerator. Simplified circuit diagram of the measurement setup is shown in Fig.~\ref{fig:Setup}. 
Standard spectroscopy techniques are used to measure a cavity resonant frequency and leak rate of $\omega_c/2\pi = 8.081~\mathrm{GHz}$ and $\kappa/2\pi = 10~\mathrm{MHz}$.
The first transition frequency and the anharmonicity of the fixed-frequency transmon qubit are $\omega_q/2\pi = 5.0353~\mathrm{GHz}$ and $\alpha = 233~\mathrm{MHz}$. The qubit-cavity coupling strength is $g/2\pi = 67.6~\mathrm{MHz}$ which results in a dispersive shift $2\chi/2\pi = 3~\mathrm{MHz}$. The relaxation time of the qubit is $T_1 = 2.8~\mu s$, while the Ramsey decay time is $T_2^* = 2~\mu s$.

The SLUG is placed in a $\mu$-metal shield and  anchored to the base plate of the dilution refrigerator. We have a microwave switch to bypass the SLUG with a through co-axial line, enabling us to calibrate the SNR of the system with the SLUG in the measurement chain. The flux and current bias of the SLUG are chosen to match the optimal bandwidth of the SLUG with the measurement resonator. The SNR improves by $7~\mathrm{dB}$ in a $20~\mathrm{MHz}$ window compared with the bare HEMT resonator. For the HEMT noise temperature of $4.1~\mathrm{K}$, this corresponds to a noise temperature of $0.8~\mathrm{K}$. 
Two isolators, providing $36~\mathrm{dB}$ of isolation, are used to protect the sample from amplifier radiation.

The improved SNR results in substantial improvement in the qubit readout. The qubit is passively initialized in the ground state and driven to the excited state with a $40~\mathrm{ns}$ Gaussian-envelope $\pi$ pulse.  After preparation, the qubit state is measured with a $2~\mu s$ microwave pulse at a frequency close to the cavity resonance.  We optimize fidelity empirically by varying system parameters, measuring fidelity with an optimal boxcar filter~\cite{Jay2007} using $40,000$ ground and excited state preparations for each set of system parameters.

We first optimize the power and frequency of the readout pulse in the linear regime\cite{Reed2010}. Increasing power increases the readout signal. However, power is limited by the breakdown of the dispersive limit $n_{\mathrm{crit}} = \Delta^2/4g^2 = 500$, above which relaxation increases dramatically. Additionally, the drive power must be below the saturation power of the SLUG.  The optimal readout power is $\bar{n}=(n_g+n_e)/2 \approx 24$ photons in the cavity and the optimal frequency is $\omega_{read} = 8.0762~\mathrm{GHz}$.  With these readout parameters, the combined preparation and readout fidelity is $91.9\%$ with $\tau = 200~\mathrm{ns}$ integration time.
From the histogram, we can define the $\mathrm{SNR}_{\mathrm{meas}} = |\mu_g - \mu_e|/(\sigma_g + \sigma_e) = 3.3$, where $\mu_g(\mu_e$) and $\sigma_g(\sigma_e$) are the means and standard deviations of the distribution of the ground(excited) state. This would correspond to a fidelity about $\mathrm{erf}(\mathrm{SNR}/\sqrt{2}) = 99.9\%$ if SNR were the only cause of fidelity loss, though clearly this cannot be achieved with imperfect preparation and lossy qubits. We can compare the SNR with the expected value for our experimental system. $\mathrm{SNR}_{theo}\approx 2\overline{sin(\theta)}\sqrt{\bar{n}\kappa\tau/(2n_{\mathrm{noise}}+1)}$, where $\overline{sin(\theta)}$ is the  
average quadrature transmission coefficient. Then we get $n_{\mathrm{noise}} = 3.2$ which agrees well with the estimate $k_BT_{\mathrm{noise}}/\hbar\omega_q = 3.3$. The improvement in the SNR allows the observation of quantum jumps in the qubit state~\cite{Vijay2011}, and this signal could be used for real-time feedback control~\cite{Riste2012FB}.

\begin{figure}
\centering
\includegraphics[scale=1]{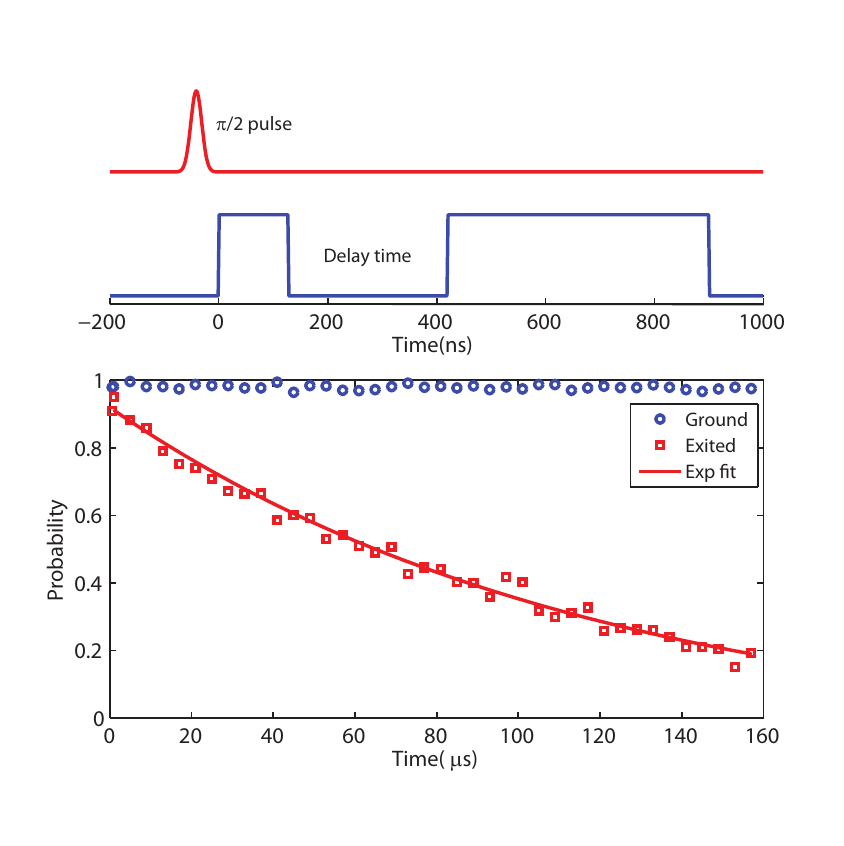}
\caption{Two pulse correlation measurement.  (a) Sequence of qubit control and readout pulses used to determine the dependence of conditional probability on the delay time between two measurements.  (b) Time evolution of conditional probability $\mathrm{P}_{g|g}(\tau)$ and $\mathrm{P}_{e|e}(\tau)$ with an exponential fit. $\mathrm{P}_{g|g}(\tau=0)\approx98.3\%$, $\mathrm{P}_{e|e}(\tau=0)\approx91.1\%$.}
{\label{fig:QND}}
\end{figure}

This qubit measurement is QND, demonstrated following the techniques of Ref~\cite{Lupascu2007}. We apply two consecutive measurement pulses after a $\pi/2$ qubit initialization pulse.
Then, we calculate the correlation between the two readout results to determine how measurement affects the qubit state. In a QND measurement with no $T_1$ processes, these two measurements will be perfectly correlated.  A delay between the two pulses are varied to see the time evolution of the correlation. We define the conditional probability $\mathrm{P}_{e|e}(\tau)(\mathrm{P}_{g|g}(\tau))$ that the qubit state is unchanged after the first measurement. We find the ground state correlation $\mathrm{P}_{g|g}(\tau=0)\approx98.3\%$[Fig.~\ref{fig:QND}], indicated a highly QND measurement.  Because the measurement is QND, it is possible to use measurement to improve preparation fidelity by initially preparing a better ground state, thus improving the combined preparation and readout fidelity.

\begin{figure}
\centering
\includegraphics[scale=1]{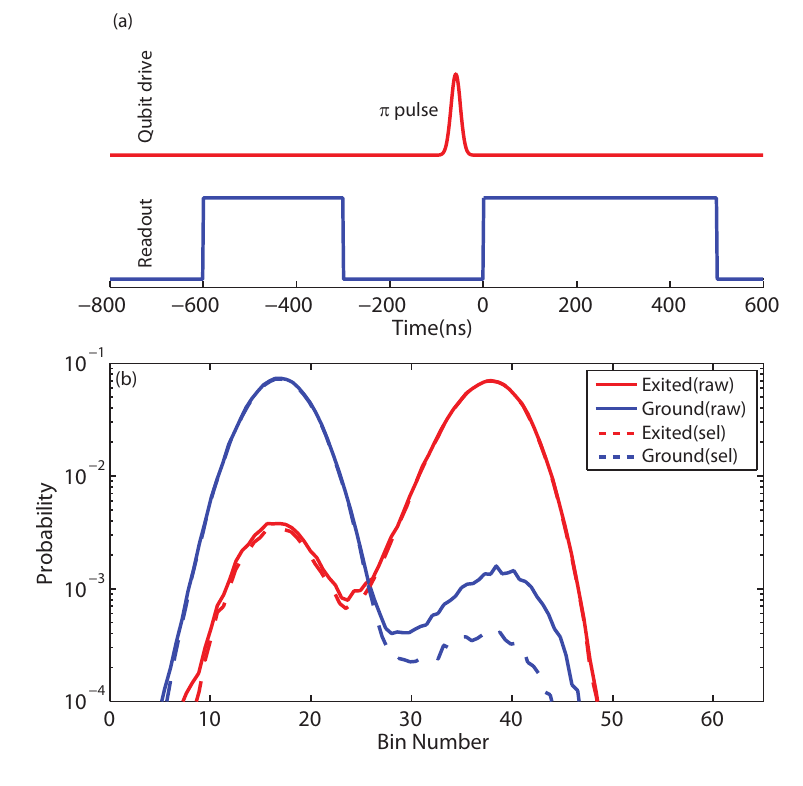}
\caption{Post-selection measurement.  (a) Sequence of qubit control and readout pulses used to suppress errors due to thermal population.  (b)Log-linear raw histogram for $~400000$ excited and ground state readout events compared with  the histogram generated after post-selection. Post-selection is based on the readout result of the first measurement pulse. By eliminating bad preparations in this way, readout error of the excited state decreases from $5.3\%$ to $4.7\%$, and the ground state from $2.8\%$ to $1.0\%$. Thus the overall fidelity increases from $91.9\%$ to $94.3\%$.}
{\label{fig:Post-selection}}
\end{figure}

The fidelity measured here is a combined preparation and readout fidelity, as either a preparation or readout error will lead to a mismatch when comparing expected and measured results.  A major source of preparation error is thermal population of the qubit, which is often hotter than the base temperature of the dilution refrigerator~\cite{Antonio2011,Barends2011}. To improve this source of preparation error, active reset methods have been proposed and realized~\cite{Riste2012FB, Geerlings2013}. In our case, the QND character of the measurement and the small overlap of state distributions make it possible to post-select true ground states before the state preparation~\cite{Johnson2012,Riste2012}. Prior to qubit manipulation, we insert a $320~\mathrm{ns}$ measurement pulse, then wait $300~\mathrm{ns}$ for the cavity to deplete of photons.  
The events where the qubit is initially determined to be in the excited state are discarded. With this technique, the fidelity rises by $2.4\%$ to $94.3\%$. We plot the histogram in log-linear mode and we can see that heating related error is greatly suppressed[Fig.~\ref{fig:Post-selection}].

After this post-selection technique, there is very little residual preparation error.  We use the randomized benchmarking(RB) protocol~\cite{RB2011} to quantify this preparation error when preparing the excited state. This RB protocol provides a reliable way to estimate the average error for a set of computational gates by applying a sequence of random gates and examining error accumulation.  In this way, the average error of a $\pi$ pulse is estimated to be $0.5\%$, which introduces a slight preparation error for excited state preparation.  The remaining infidelity is primarily a measurement error due to relaxation during the measurement pulse.

To conclude, we have implemented dispersive readout of a transmon qubit with high single-shot fidelity. A low noise SLUG amplifier and other optimization techniques are shown to significantly improve the readout SNR, providing $94.3\%$ combined readout and preparation fidelity with a highly QND measurement. A similar result is recently reported~\cite{Hover2013}. As the SLUG requires only DC bias, has high dynamic range, and can easily be isolated from the qubit, it provides a possible alternative to parametric amplifiers.  Fidelities larger than $98\%$ could be possible in devices with $T_1>10~\mu s$
~\cite{Chang2013,Paik2011,Chad2012}. 

This work was supported by IARPA under Contract W911NF-10-1-0324

%

\end{document}